\documentclass[10pt]{article}
\usepackage[latin1]{inputenc}
\usepackage{graphics}
\usepackage{amssymb}
\usepackage{amsfonts}
\usepackage[thmmarks]{ntheorem}

\def\CC{{\mathbb C}}

\def\bra{\langle}
\def\ket{\rangle}

\def\bea{\begin{eqnarray}}
\def\eea{\end{eqnarray}}

\def\be{\begin{equation}}
\def\ee{\end{equation}}

\newtheorem{theorem}{Theorem}

\newtheorem{lemma}{Lemma}

\newtheorem{cor}{Corollary}
 
\theoremstyle{nonumberplain}
\theorembodyfont{\normalfont}
\theoremseparator{:}
\theoremsymbol{$\P$}

\newcommand\transp[1]{\hspace{0.4mm}{}^t\hspace{-0.6mm}#1}
\begin{document}

\title{A remark on the mathematics of the seesaw mechanism}
\author{Fabien Besnard\footnote{Pôle de recherche M.L. Paris, EPF, 3 bis rue Lakanal, F-92330 Sceaux. \goodbreak fabien.besnard@epf.fr}}
\maketitle
\begin{abstract}
To demonstrate that matrices of seesaw type lead to a hieararchy in the neutrino masses, i.e. that there is a large gap in the singular spectrum of these matrices, one generally uses an approximate block-diagonalization procedure. In this note we show that no approximation is required to prove this gap property if the Courant-Fisher-Weyl theorem is used instead.   This simple  observation might not be original, however it does not seem to show up in the literature. We also sketch the proof of additional inequalities for the singular values of matrices of seesaw type.
\end{abstract}

\section{Introduction}
The terms in the Standard Model Lagrangian\footnote{extended with right-handed neutrinos}  giving mass to the neutrinos can be gathered in a matrix of the general form (see for instance \cite{ms})

\be
M_\nu=\pmatrix{m_L&\transp{m_D}\cr m_D&M_R}
\ee

which has to be symmetric and where each entry is a complex $3\times 3$ matrix acting on the generation (or flavor) space. The requirement of renormalizability gives the further contraint that $m_L=0$. (Note that the Noncommutative Geometry approach to the Standard Model naturally predicts a matrix of this type with $m_L=0$ without any consideration of renormalizability, see \cite{BF}, \cite{BBB}).

The neutrino masses are the singular values of $M_\nu$, that is to say the eigenvalues of the positive definite matrix $\sqrt{M_\nu^*M_\nu}$, where the star means matrix adjoint.

To explain the smallness of the observed neutrino masses it is generally argued that if $m_D$ is small with respect to $M_R$, then the singular values of $M_\nu$ split into two families: one very small, and one large (of the order of $M_R$). This is the seesaw mechanism. It is easy to show explicitly for one generation, since in that case $M_\nu$ is a $2\times 2$ matrix. It is then found that the smallest singular value $m_\nu^1$ of $M_\nu$ satisfies 

\be
{m_\nu^1\over m_R}\approx \left({m_D\over m_R}\right)^2
\ee

at second order in ${m_D\over m_R}$  while the largest singular value $m_\nu^2$ is approximately equal to $m_R$. With just a little more work one finds  in fact that (with $m_D$ and $m_R$ real)

\bea
{m_\nu^1\over m_R}&\le& \left({m_D\over m_R}\right)^2 \cr
 {m_\nu^2\over m_R} &>& 1\label{onegen}
\eea

We call this ``the gap property''. For three generations though, $M_\nu$ cannot be diagonalized by an analytical formula and one appeals to an approximate block diagonalization (see for instance \cite{LOS} or the appendix of \cite{SS}), that is to say that $M_\nu$ is brought to block-diagonal form thanks to approximately unitary matrices. One can then prove the gap property up to higher order terms.

Our purpose here is just to make the simple observation that  if one uses the Courant-Fischer-Weyl theorem then no approximation is needed to prove the gap property for the singular values     for an arbitrary number of generations, in the form of exact inequalities like (\ref{onegen}).

More precisely, let $M_\nu$ be the symmetric complex $2n\times 2n$ matrix

\be
M_\nu=\pmatrix{0&\transp{m_D}\cr m_D&M_R}\label{sst}
\ee

where $n$ is the number of generations. Let $m_\nu^1,\ldots,m_\nu^{2n}$ be the singular values of $M_\nu$ written in ascending order. (Hence if $n=3$, $m_\nu^1\le m_\nu^2\le m_\nu^3$ are the masses of the $3$ light neutrinos at tree level, and $m_\nu^4\le m_\nu^5\le m_\nu^6$ are the masses of the $3$ heavy ones.) Let $m_D^1\le \ldots\le m_D^n$ be the singular values of $m_D$ (Dirac masses) and $m_R^1\le \ldots\le m_R^n$ be the singular values of $m_R$ (Majorana masses). We further suppose that $m_D$ is not singular (hence $m_D^1>0$) and that $m_D^n<m_R^1$. Then we will show that

\bea
m_\nu^n & \le & {(m_D^n)^2\over\sqrt{(m_D^n)^2+(m_R^1)^2}}\cr
m_\nu^{n+1}&\ge&\sqrt{(m_R^1)^2+(m_D^1)^2}\label{estimate}
\eea

which immediately entails

\be
m_\nu^1\le\ldots\le m_\nu^n<{(m_D^n)^2\over m_R^1}<m_R^1<m_\nu^{n+1}\le\ldots\le m_\nu^{2n}\label{gap} 
\ee

which directly generalizes (\ref{onegen}) to $n$ generations. 

It is our hope that these exact formulas can be of some use to physicists. Here are some motivations for this hope:

\begin{enumerate}
\item In the usual approximate method one reasons on the order of magnitude of the entries of the matrices $m_D$ and $m_R$. However a matrix with large entries can have very small (even vanishing) singular values. The method exposed here could be used to see what are the most general relations to be expected among the singular values without getting our hands dirty by delving into the algebraic relations satisfied by the matrices.
\item This method is also fairly general (in particular it does not depend on any ansatz as the approximate block-diagonalization does). It might be useful in other contexts where the ratio of the entries of $m_D$ on those of $m_R$, though smaller than one, is not so small as to completely neglect all the multiplicative constants introduced in every step of the approximation.
\end{enumerate}

The paper is organized as follows: in section 2 we  recall the necessary mathematical background keeping it to the minimum required to prove the gap property in section 3. In section 4 we sketch the proof of additional  inequalities for the singular values thanks to immediate generalizations of the formulas in section 2.




\section{Min-max theorem and matrix inequalities}

We recall here the following theorem.

\begin{theorem} (Courant-Fischer-Weyl min-max theorem) Let $M$ be a self-adoint $N\times N$ matrix with eigenvalues $m_1\le \ldots\le m_N$. Then: 
$$m_k=\min_W(\max\{\bra MX,X\ket|X\in W,\|X\|=1\})$$
where $W$ runs over all vector subspaces of $\CC^N$ of dimension $k$, and

$$m_k=\max_W(\min\{\bra MX,X\ket|X\in W,\|X\|=1\})$$
where $W$ runs over all vector subspaces of $\CC^N$ of dimension $N-k+1$.
\end{theorem}

This yields the following well-known corollary that we will need. For a self-adjoint matrix $M$ let us write $\min(M)$ for the smallest eigenvalue of $M$.

\begin{cor}\label{cor1} Let $A,B$ be two self-adjoint $N\times N$ matrices. Then 
$$\min(A+B)\ge \min(A)+\min(B)$$
\end{cor}


Thanks to the min-max theorem one can also easily show the following interlacing property (called the Cauchy interlacing theorem): let $Q$ be a submatrix of $M$ obtained by orthogonal projection on a vector subspace generated by $n$ basis vectors. Let $q_1\le \ldots\le q_n$ be the eigenvalues of $Q$. Then

\be
m_k\le q_k\le m_{N-n+k}\label{cauchy}
\ee

for every $k\le n$. In the main part of this paper we will only need the special case where $N=2n$ and $k=1$, yielding

\be
\min(Q)\le m_{n+1}\label{inter}
\ee

We will also need the following  lemma:

\begin{lemma}\label{lemma1}  For any $A\ge 0$ and any $B\in M_n(\CC)$ one has

$$\min(B^*AB)\ge \min(A)\min(B^*B)$$
\end{lemma}

We prove the lemma. It is obvious when $B$ is singular. We then suppose that it is not. Let $X$ be a unit vector. We have:

\bea
\bra B^*ABX,X\ket&=&\bra ABX,BX\ket\cr
&=&\bra A{BX\over\|BX\|},{BX\over \|BX\|}\ket\|BX\|^2\cr
&\ge& \min(A)\|BX\|^2,\mbox{ by the min-max theorem}
\eea

Now $\|BX\|^2=\bra B^*BX,X\ket\ge \min(B^*B)$ also by the min-max theorem. Since $\min(A)\ge 0$ one gets $\bra B^*ABX,X\ket\ge \min(A)\min(B^*B)$, frow which the results follows using the min-max theorem again.

\section{Singular value estimates for matrices of seesaw type}

From (\ref{sst}) we compute

\be
M_\nu M_\nu^*=\pmatrix{?&?\cr ?&{m_D}{m_D}^*+M_R M_R^*}\label{aetoilea}
\ee

where the question marks stand for matrices we do not care about. We call $Q={m_D}{m_D}^*+M_R M_R^*$.  From (\ref{inter}) we get:

\be
\min(Q)\le m_{n+1} 
\ee

where $m_{n+1}=(m_\nu^{n+1})^2$. But $\min(Q)\ge \min(M_RM_R^*)+\min({m_D}{m_D}^*)=(m_R^1)^2+(m_D^1)^2$  which yields the second part of (\ref{estimate}).
\smallbreak

To prove the first part we first need to write down the inverse of $M_\nu$. There exists a general formula for inverting $2\times 2$ block matrices. Here we can check by direct computation that 

\be
M_\nu^{-1}=\pmatrix{-m_D^{-1}M_R\transp{m_D}^{-1}&m_D^{-1}\cr \transp{m_D}^{-1}&0}
\ee

We then see that $(M_\nu^*M_\nu)^{-1}=\pmatrix{X&?\cr ?&?}$, where

\be
X=m_D^{-1}M_R\transp{m_D}^{-1}(\transp{m_D}^{-1})^*M_R^*(m_D^{-1})^*+(m_D^*m_D)^{-1}\label{xx}
\ee

Using (\ref{inter}) again we obtain

$$\min(X)\le m_{n+1}$$

where this time $m_{n+1}$ is $n+1$-th largest eigenvalue of $(M_\nu^*M_\nu)^{-1}$, that is to say $m_{n+1}=(m_\nu^n)^{-2}$. Hence 

$$\min(m_D^{-1}M_R\transp{m_D}^{-1}(\transp{m_D}^{-1})^*M_R^*(m_D^{-1})^*)+\min((m_D^*m_D)^{-1})\le {1\over (m_\nu^n)^2}$$

Now using the lemma twice we obtain

\bea
\min(m_D^{-1}M_R\transp{m_D}^{-1}(\transp{m_D}^{-1})^*M_R^*(m_D^{-1})^*)&\ge&\min(M_R\transp{m_D}^{-1}(\transp{m_D}^{-1})^*M_R^*)\min(m_D^{-1}(m_D^{-1})^*)\cr
&\ge&\min(\transp{(m_Dm_D^*)}^{-1})\min(M_RM_R^*)\min((m_D^*m_D)^{-1})\cr
&\ge&\min((m_Dm_D^*)^{-1})\min(M_RM_R^*)\min((m_D^*m_D)^{-1})\cr
&\ge&{(m_R^1)^2\over (m_D^n)^4}
\eea

We thus have
$${(m_R^1)^2\over (m_D^n)^4}+{1\over(m_D^n)^2}\le {1\over(m_\nu^n)^2}$$

which easily yields the first part of (\ref{estimate}).




\section{Additional inequalities}
We now sketch the proof of the following inequalities:

\be
m_\nu^{n+k}\ge \sqrt{(m_D^1)^2+(m_R^k)^2}\label{ineq1}
\ee
for $k=1,\ldots,n$ and 
\be
m_\nu^{j}\le {m_D^n m_D^{j}\over \sqrt{(m_D^{j})^2+(m_R^1)^2}}\label{ineq2}
\ee
for $j=1,\ldots,n$.

For this we will need to strengthen corollary \ref{cor1} and lemma \ref{lemma1}. The first strengthening is given by  Weyl's inequalities: if $A$ and $B$ are hermitian $n\times n$ matrices, and $C=A+B$, then for $1\le k\le n$ one has

\be
a_k+b_1\le c_k\le a_k+b_n\label{weyl}
\ee

where $c_1\le\ldots\le c_k$, $a_1\le\ldots\le a_n$ and $b_1\le\ldots\le b_n$ are the eigenvalues of $A,B$ and $C$.

 As for lemma \ref{lemma1}, we can extend it in the following way. 

\begin{lemma}\label{lemma2}  For any $A\ge 0$,  $B\in M_n(\CC)$ let $C=B^*AB$. Then one has (with the same notations as above)

$$c_k\ge a_k\min(B^*B)$$
\end{lemma}

The proof of this lemma follows the same line as the one of lemma \ref{lemma1}. Suppose $B$ is non singular and let $W$ be a subspace of $\CC^n$ of dimension $k$. Then $BW$ has dimension $k$ and must intersect the orthogonal of the subspace generated by the $k-1$ eigenvector of $A$ corresponding to $a_1,\ldots,a_{k-1}$. Hence $\bra ABX,BX\ket\ge a_k\|BX\|^2$ on $W$. The result follows from the minmax theorem.

The inequalities (\ref{ineq1}) and (\ref{ineq2}) can then be  proven by the same techniques as in the previous section.

\end{document}